# Probing quantum walks through coherent control of high-dimensionally entangled photons

Poolad Imany,[1,2,†] Navin B. Lingaraju,[1,2] Mohammed S. Alshaykh,[1,2] Daniel E. Leaird,[1,2] Andrew M. Weiner[1,2]

[1]School of Electrical and Computer Engineering, Purdue University, West Lafayette, IN, USA.

[2]Purdue Quantum Science and Engineering Institute, Purdue University, West Lafayette, IN, USA.

† pimany@purdue.edu

**Abstract: Quantum walks in atomic systems, owing to their continuous nature, are especially well-suited for the simulation of many-body physics and can potentially offer an exponential speedup in solving certain black box problems. Photonics offers an alternate route to simulating such nonclassical behavior in a more robust platform. However, in photonic implementations to date, an increase to the depth of a continuous quantum walk requires modifying the footprint of the system. Here we report continuous walks of a two-photon quantum frequency comb with entanglement across multiple dimensions. The coupling between frequency modes is mediated by electro-optic phase modulation, which makes the evolution of the state completely tunable over a continuous range. With arbitrary control of the phase across different modes, we demonstrate a rich variety of behavior: from walks exhibiting ballistic transport or strong energy confinement, to subspaces featuring bosonic or fermionic character. We also explore the role of entanglement dimensionality and demonstrate biphoton energy bound states, which are only possible with multilevel entanglement. This suggests the potential for such walks to quantify entanglement in high-dimensional systems.**

A quantum particle can exist in a superposition of paths, or modes, and interference between the probability amplitudes of these outcomes results in phenomena unique to random walks of quantum systems (*1–3*) – enhanced propagation, otherwise called ballistic transport (*4*), or Anderson localization, where the wavefunction becomes confined in a disordered system (*5, 6*). Quantum walks of two or more particles can exhibit nonclassical phenomena such as bunching or antibunching for bosons and fermions, respectively (*7*). The complex dynamics exhibited in such walks cannot be explained by classical models and, therefore, can serve as a probe of entanglement or interactions between particles (*1, 7–11*). Owing to the variety of nonclassical behavior they can exhibit, quantum walks have the potential to provide a dramatic speed up in certain computational tasks like physical database searches (*12*) and tests of graph isomorphism (*13*).

Quantum walks come in two flavors – continuous and discrete (*14*). Continuous quantum walks have been observed in atomic systems (*1*), where the depth of the walk is determined by the evolution time of the quantum state. The continuous nature of these quantum walks allows one to explore the full range of parameter space, which makes such walks especially well-suited to simulating Hamiltonian dynamics and solving certain black box problems exponentially faster (*15*). To achieve even comparable performance with discrete quantum walks requires additional system complexity, primarily through an extra degree of freedom (*14*). Quantum walks have also been implemented in photonic systems as they offer a more robust platform in terms of decoherence and room-temperature operation (*2, 3, 9, 16*). However, photonic quantum walks demonstrated to date suffer from the drawback that their circuit depth can only be incremented in discrete steps and by physically altering the footprint of the system.

Here we report continuous quantum walks of photon pairs entangled across multiple dimensions (*8, 17*). The depth of our quantum circuit is fully tunable without requiring any change to the scale of the system. With arbitrary control of the phase across different modes, we demonstrate walks exhibiting ballistic transport or strong energy confinement, as well as subspaces featuring bosonic or fermionic character. We also explore the role of entanglement dimensionality and demonstrate biphoton energy bound states, which are only possible with entanglement across multiple dimensions. This sensitivity to multilevel entanglement suggests the potential for such walks to quantify entanglement in high-dimensional systems.

A photon can "walk" along different modes in any one of its many degrees of freedom (DoFs), whether it be time (*2, 3*), path (*8, 16, 18*), orbital angular momentum (*19*), or frequency (*20*). All that is required to observe such behavior is the presence of coupling between different modes in the particular DoF. In the case of quantum walks in the frequency domain, this coupling is mediated by a periodic (temporal) modulation of the waveguide refractive index. Such coupling, or mode-splitting, can be realized in electro-optic phase modulators driven with a single sinusoidal radiofrequency (RF) tone. The effect of this perturbation is that the wavefunction of a photon traversing the waveguide picks up a factor of $e^{i\delta \cos \omega_m t}$. Here, $\delta$ corresponds to the strength, or depth, of the modulating RF field and $\omega_m$ denotes the frequency of modulation. From the perspective of the frequency domain, phase modulation scatters a single frequency into a comb-like spectrum where adjacent frequency modes are separated by $\omega_m$ in frequency [Fig. 1a]. The amplitude of a comb line a distance $n\omega_m$ away from the original frequency is given by $n^{th}$-order Bessel function $J_n(\delta)$. The power of this process, in the context of quantum walks, is immediately apparent – unlike the case where a particle can move only one step to the left or to the right, here a photon is projected into a multitude of modes in a single step; acting, in effect, as a multi-port beam splitter (*21, 22*). In analogy to quantum walks based on path encoding, the depth of such a frequency-domain quantum walk can be incremented by simply cascading one modulator after another. However, an added strength of the frequency domain approach is that a cascade of $n$ identical phase modulators is equivalent to increasing the strength of the modulating RF field in a *single* phase modulator by a factor of $n$. In other words, the depth of a quantum walk in the frequency domain can tuned over a *continuous* range – a feature yet to demonstrated with photons in any other DoF – by simply modifying the strength of the modulating field.

In Fig. 1a, we show results from a quantum walk of a single photon starting out in a single frequency. As the strength of the modulating RF field increases (δ), the extent to which the input mode spreads to outer frequencies also increases. Since modulation strength δ is equivalent to circuit depth in our approach, we recorded the photon distribution across various output modes for different values of δ. The "rabbit ears" observed in this distribution (Fig. 1a) signifies the presence of ballistic energy transport, which is a signature of random walks with quantum systems. In particular, the standard deviation of the output photon distribution grows linearly with $\delta$ (*23*). For a classical random walk, transport to neighboring modes is not nearly as fast and its standard deviation grows only as $\sqrt{\delta}$ (*23*). The presence of entanglement, particularly high-dimensional entanglement, makes it possible to observe a richer variety of behavior than is possible with just a photon in a single frequency mode. To explore the effect of entanglement, we studied the evolution of two-photon quantum frequency combs, commonly referred to as biphoton frequency combs (BFCs) [Fig. 1b], in a quantum walk.

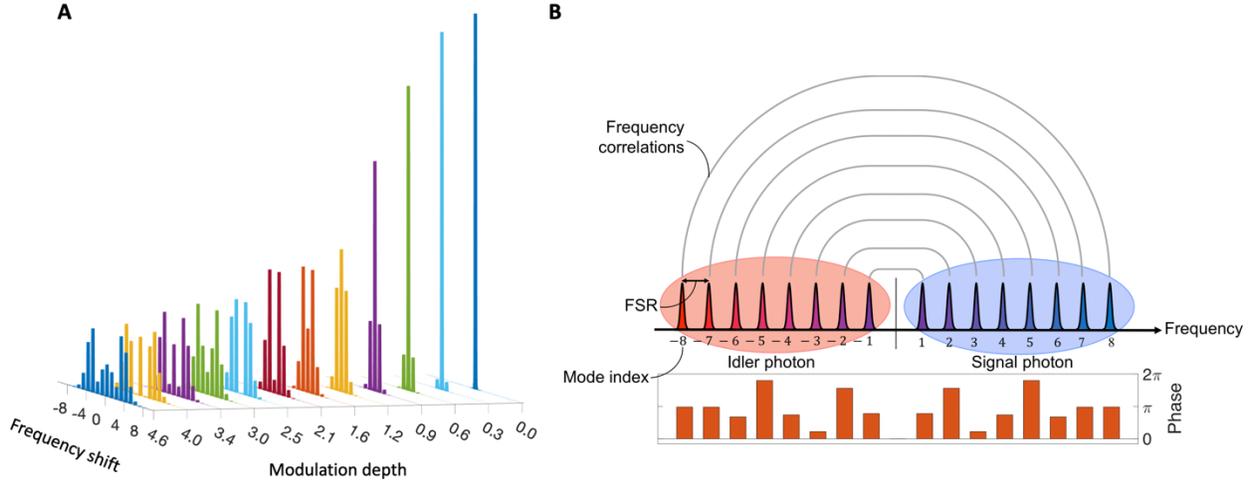

**Fig. 1. Frequency domain mode splitting and biphoton frequency comb.** (**A**) Experimental data showing the effect of phase modulation on a single frequency mode for various modulation depths. These spectra were acquired with classical light but also serve as an illustration of how each single mode, even in the quantum regime, is scattered by a phase modulator. The modulation speed was chosen to match the mode separation in our quantum source and, therefore, frequency shifts are presented in terms of the BFC mode index. (**B**) Illustration of a biphoton frequency comb (BFC) in frequency space. Each photon of the entangled pair is in a superposition of eight distinct frequency modes with pairwise correlations about center of the biphoton spectrum. This two-photon state has the form $|\psi\rangle = 1/\sqrt{8} \sum_{m=1}^{8} |m, -m\rangle_{SI}$ and gray lines in the figure highlight correlations between each frequency pair $|m, -m\rangle_{SI}$. A pulse shaper is used to manipulate the phase on each frequency mode prior to the quantum walk.

BFCs have been generated directly in on-chip optical microresonators (*24–27*) or carved from continuous down-conversion spectra (*28*). For results reported here, BFCs were generated by the latter approach as it allows flexibility in the choice of comb linewidth ($\Delta\lambda$) and free spectral range (FSR). Broadband time-energy entangled photons (~5 THz) were generated by type-0 down-conversion in a periodically-poled lithium niobate (PPLN) waveguide. In this process, a pump photon from a continuous wave laser (~775 nm) is converted into a pair of daughter photons in telecommunications band (~1520-1580 nm). As energy is conserved in this process, the frequencies of the daughter photons must add up to that of the pump photon. In other words, the two photons in an entangled pair are anticorrelated in frequency. This two-photon spectrum is carved into a BFC using a Fourier transform pulse shaper. The 3dB linewidth of each frequency mode is set to 9 GHz and is limited by the resolution of the pulse shaper. To ensure minimal crosstalk between adjacent modes, the FSR of the BFC is chosen to be 25 GHz.

The pulse shaper is used to not only manipulate the amplitude of the biphoton spectrum, but also its phase prior to any quantum walk. In particular, the spectral phase can be set to vary continuously or to make discrete jumps from one mode to the next. Once the desired state has been prepared, it is sent to a phase modulator to implement the mode-mixing operations that give rise to a quantum walk. Our modulator is driven with a 25 GHz sinusoidal RF waveform, identical to the FSR of the BFC, with tunable power over a continuous range. Downstream of the phase modulator is a second pulse shaper, which is used to select a pair of output frequencies and routes each one to a superconducting nanowire single photon detector (SNSPD). Two-photon events are identified by correlations in their arrival time and this data is used to construct a measurement of the joint spectral intensity (JSI) of the BFC, which illustrates the effects of a quantum walk in energy space.

As noted earlier, one hallmark of quantum walk is the observation of ballistic energy transport between the quantum state and system through which it walks. We demonstrate such behavior for the case of a BFC entangled across eight dimensions that has the form $|\psi\rangle = 1/\sqrt{8} \sum_{m=1}^{8} |m, -m\rangle_{SI}$, where S and I denote the signal (high-frequency) and idler (low-frequency) photons, respectively. The JSI of this state, i.e., in the absence of any phase modulation, is completely anticorrelated in frequency [Fig. 2a]. In Fig. 2b we show the JSI of this state after traversing a phase modulator driven to a depth $\delta = 4.6$ (see supplementary material for the JSI measurements of various modulation depths). Experimental data clearly show migration of the state away from the original JSI, which matches results expected from theory (see supplementary material). Transport to the top right triangle of the JSI plot corresponds to instances where the phase modulator transfers energy to the biphoton. The converse, when the biphoton transfers energy to the modulator, manifests as transport to the lower left triangle of the JSI. In other words, what we observe in frequency space is photon bunching – a characteristic of bosons. Energy transfer between the quantum circuit and the two photons, as a function of modulation depth, is shown in Fig. 2d. The standard deviation of the energy transfer, as in the single photon case, is linear with $\delta$ [Fig. 3a]. However, this linear rate of energy transfer is roughly twice as fast in the case of entangled photons [Fig. 3a]. Such enhanced energy transport was previously demonstrated with high-dimensional, path-entangled photon pairs (*8*).

We also break new ground in photonic quantum walks by demonstrating the opposite of ballistic energy transport – strong energy confinement of the biphoton irrespective of modulation or circuit depth. To achieve this, we manipulate the spectral phase on the BFC to create $|\psi\rangle = 1/\sqrt{8} \sum_{m=1}^{8} e^{im\pi} |m, -m\rangle_{SI}$, i.e., a state in which adjacent modes have a $\pi$ phase with respect to one another. This operation can be viewed as a linear spectral phase shift, which is equivalent to delaying one photon, with respect to its entangled counterpart, by half the modulation period. As a result of this delay, photons in an entangled pair acquire *equal and opposite* frequency shifts. This is clearly illustrated in the JSI measurement after a quantum walk [Fig. 2c]. Frequency (energy) correlations remain confined to the constant energy axis, i.e., antidiagonal of the JSI measurement. As the circuit depth is increased, frequency correlations merely propagate outward along the antidiagonal to include new combinations of high and low photon energies. While the energy of the energy of the two-photon state remains constant, individual photons in a pair do gain and lose energy as they progress through the circuit. However, this energy gain or loss is correlated within a photon pair. If the idler gains some energy, the signal loses that same amount of energy, i.e., the photons antibunch in energy space. Antibunching of parties is a distinctive feature of fermionic quantum walks and has also been demonstrated in other photonic platforms (*7*).

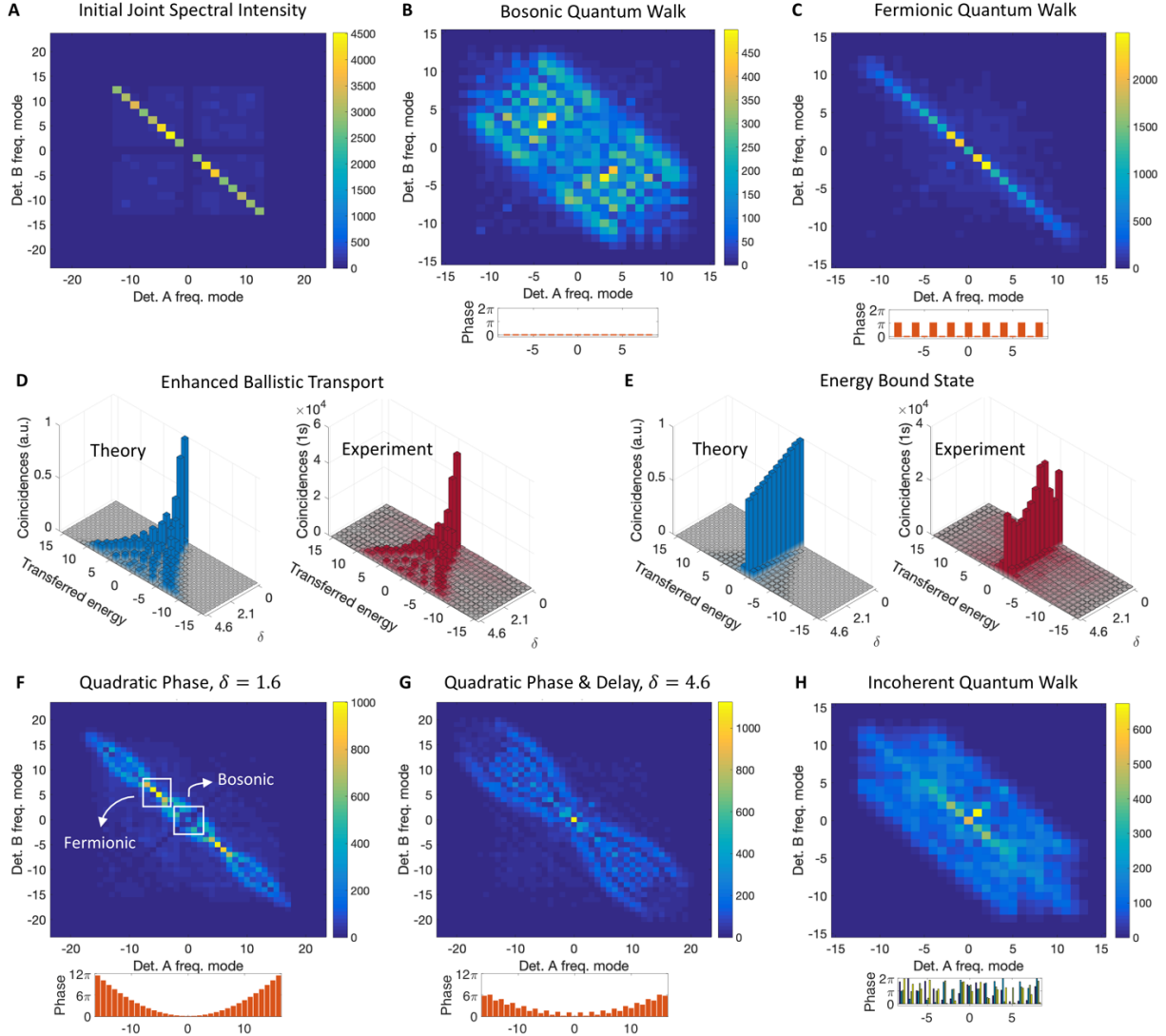

**Fig. 2. Two-photon quantum walks.** (**A**) A measurement of the joint spectral intensity (JSI) for an eight-dimensional biphoton frequency comb (BFC) prior to the quantum walk. Coincidences are observed for mode pairs $m, -m$, which are anticorrelated in frequency. The JSI is symmetric about the diagonal as any two-photon event $|i,j\rangle_{AB}$ is equivalent to its mirror $|j,i\rangle_{AB}$. Diagonal terms $|i,i\rangle_{AB}$ were measured by splitting frequency mode $i$ between detector channels A and B. The acquisition time for diagonal elements was twice as long since there is a 50% probability that both photons end up at the same detector and, consequently, fail to register coincidences. (**B**) JSI for a BFC after a quantum walk for the case when no additional phase is applied prior to the walk ($|\psi\rangle = 1/\sqrt{8} \sum_{m=1}^{8} |m,-m\rangle_{SI}$). This results in two-photon events where mode indices move in the same direction, i.e., we have bunching of photons in energy space. (**C**) Antibunching (mode indices of two-photon events move in opposite directions) is observed when adjacent modes start out with a $\pi$ phase difference relative to one another ($|\psi\rangle = 1/\sqrt{8} \sum_{m=1}^{8} e^{im\pi}|m,-m\rangle_{SI}$). (**D, E**) Energy transferred from the phase modulator to the total biphoton state. The bosonic-like quantum walk exhibits ballistic energy transport and we see strong energy confinement for the fermionic-like walk. The JSI for each step, or each increment to the modulation depth, is shown in supplementary material. Energy transfer in these plots is presented in units of $h\nu$ where $\nu = 25$ GHz ($h\nu = 1.656 \times 10^{-23}$ J). The variation in the coincidence rates shown in (**E**) is due to fluctuations of the photon flux in our entangled pair source. (**F**) The application of quadratic spectral phase (equivalent to 1800 m of single mode fiber) to a 16-dimensional BFC results in energy subspaces with either bosonic or fermionic character. The lower applied modulation depth compared to previous JSIs results in smaller diagonal spreading in the bosonic subspaces. (**G**) Results for a walk similar to that in (**F**), but with higher modulation depth and smaller quadratic phase (equivalent to 900 m of single mode fiber). An additional linear phase was applied to ensure energy confinement at the center of the JSI, with the transition from fermionic to bosonic character occurring further away along the antidiagonal. (**H**) Quantum walk for a mixed state that has the same initial JSI as the state in (**A**). There is no indication of either ballistic energy transport or energy confinement, pointing to a clear distinction between correlated and entangled quantum walks. All the JSI elements are coincidences measured in 1 second.

However, as our photons walk in frequency (energy) space, antibunching in a high-dimensional state has the effect of preserving the total energy of the biphoton, thereby forming a *biphoton energy bound state*. This can be seen quite clearly in Fig. 2e where we present data for energy transfer between the phase modulator and the biphoton for different modulation depths.

One decided advantage of our platform is that we can implement continuous quantum walks in a photonic platform. However, with arbitrary control over the phase across different modes, we are able to also simulate nonclassical behavior heretofore unrealized with other DoFs. By preparing BFCs with quadratic spectral phase, we are able to observe remarkable features in two-photon correlations – distinct subspaces featuring bosonic or fermionic character. Figs. 2f, g show results from such a quantum walk with a 16-dimensional entangled state. In increasing the number frequency modes across which the photons are entangled (16 compared 8 in preceding experiments), we are able to clearly delineate regions exhibiting bunching from those exhibiting antibunching [Figs. 2f, g].

The critical role played by spectral phase hints at strong differences between quantum walks featuring coherent superpositions of multiple frequency pairs $|m, -m\rangle_{SI}$ as compared to mixtures of those same frequency pairs. Although both states have identical frequency correlations, in the case of the latter the relative phase between any two basis states ($|m, -m\rangle_{SI}$ and $|m', -m'\rangle_{SI}$ for $m \neq m'$) is completely random. To simulate this random phase, we can construct a JSI measurement of the mixed state by simply adding together JSI measurements for individual frequency pairs $|m, -m\rangle_{SI}$ for $m = 1, ..., 8$ traversing the circuit (see supplementary material). A clear effect of incoherence is that frequency correlations are smeared out across energy space without any sharp or well-defined features [Fig. 2h]. The simulation of bosonic, fermionic, anyonic, and incoherent quantum walks for the limit of high modulation depth ($\delta = 200$) is shown in the supplementary material, showing clear bunching and antibunching in the bosonic and fermionic case, respectively.

This comparison allows us to observe qualitative difference between systems with different amounts of entanglement. For a more quantitative picture, we reexamine the case of a quantum walk for an energy bound state with a particular emphasis on entanglement dimensionality. While this process was presented from the perspective of energy exchange between the phase modulator and transiting photons, it is also easily understood from a time-domain illustration of the process, and by exploring the effect of phase modulation on the time correlation function of entangled photons (*29*). In Fig. 3b,c, the strength of the modulating RF waveform is shown (in blue) as a function of time. The signal photon, which can arrive at the modulator at any time owing to the random nature of the generation process, is designated by a black arrow. In Fig. 3b,c we only show one possible arrival time. Here, for example, the signal reaches the modulator when the phase of the RF waveform is $\pi/6$. While the idler photon also reaches the modulator at a random time, its arrival is highly correlated with that of the signal photon. This correlation is characterized by a distribution of possible values for the delay between signal and idler. The distribution, in delay space, is given by the Fourier transform of the complex biphoton spectrum (*30*). Consequently, for a narrowband biphoton spectrum with entanglement across a limited number of dimensions ["Idler (8-D)" in Figs. 3b], there is wide range of possible values for the delay between signal and idler. As the entanglement dimensionality of the state increases, i.e., as the biphoton spectrum gets broader, the distribution of possible delays gets narrower ["Idler (64-D)" in Figs. 3b]. The discretization of the biphoton spectrum in frequency space, owing to its comb-like structure, results in a distribution of relative arrival times that repeats at multiples of the inverse comb FSR. Since the spacing between comb lines matches the frequency of the RF waveform, this repetition of the distribution in arrival times occurs at integer multiples modulation period. The net effect is that both photons "see" nearly the same phase modulation slope ($d\phi/dt$), which means they experience the same instantaneous frequency shift (*31*).

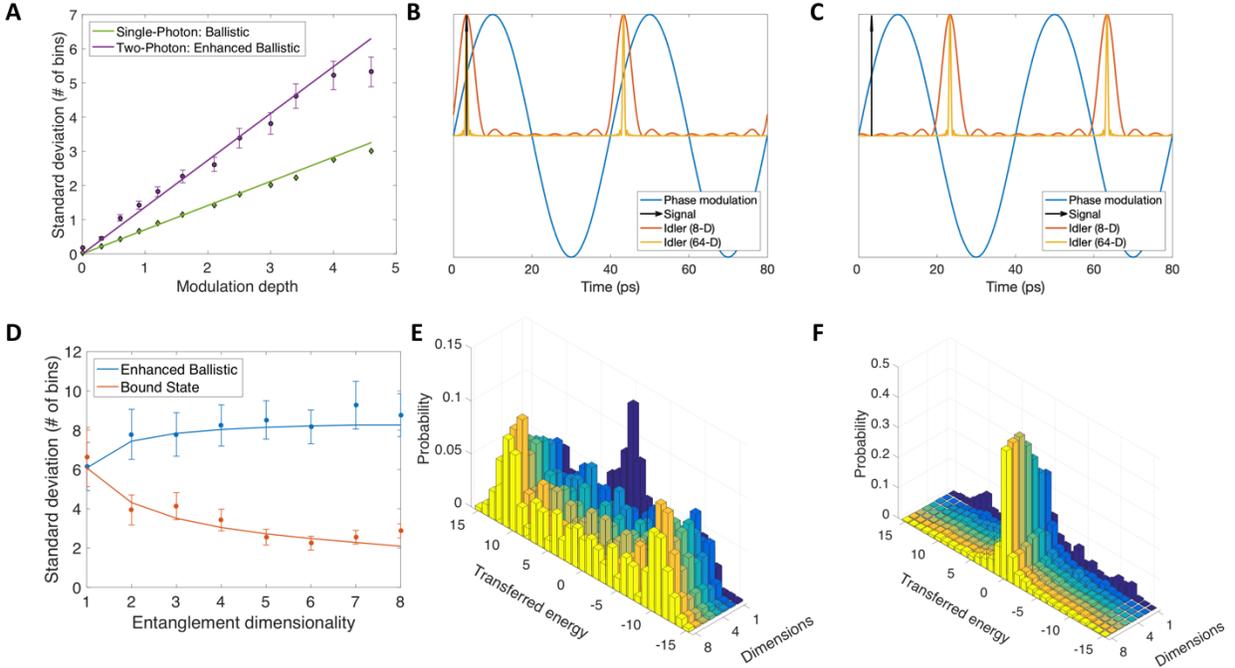

**Fig. 3. Effects of entanglement dimensionality on quantum walks. (A)** Standard deviation of single- and two-photon energy transport. In particular, the two-photon case considers an eight-dimensional, maximally entangled photon pair. Experimental data (purple and green markers) is plotted alongside results expected from theory (solid lines). Plot points are extracted from results Figs. 1a, 2d. The standard deviation grows linearly with modulation index in both cases. However, the slope is roughly twice as steep for the two-photon case. **(B, C)** Time-domain illustration of processes in a walk. The strength of the modulating RF waveform is shown in blue. A signal photon (black arrow) arrives at the modulator at random times owing to the nature of the photon pair generation process. However, the arrival time of the idler photon is highly correlated with that of the signal photon and is characterized by a distribution of joint arrival times that repeats at multiples of the BFC free spectral range. Since the spacing between the comb lines in the BFC is set to match the modulation frequency, the period of pulse-like features in the BFC time correlation function matches the period of the driving RF waveform. As the number of frequency modes across which the photons are entangled increases, the tighter the distribution of arrival times becomes. **(B)** In the case where no phase is applied to the initial state, relative timing between signal and idler photons reduces to an integer multiple of the modulation period. Consequently, both photons in a pair experience the same frequency shift, which results in enhanced ballistic energy transport. **(C)** Conversely, when there is a relative $\pi$ phase difference between adjacent modes, the relative timing between signal and idler photons is instead centered at an odd half-integer multiple of the modulation period. The net effect is that photons in a pair experience equal, but opposite, frequency shifts, forming a biphoton energy bound state. **(D)** Standard deviation of the energy transfer (output mode) distribution as a function of entanglement dimensionality for the case of enhanced ballistic transport and energy confinement when $\delta = 6.1$. Theoretical predictions are represented by solid lines and the markers correspond to experimental data extracted from **(E, F)**. In **(A, D)**, the standard deviation is computed after background subtraction (coincidence-to-accidental ratio ~ 50) and the error bars are calculated assuming Poissonian statistics. The error bars for single-photon energy transport in **(A)** are not shown since the experiment was carried out using classical light. **(E, F)** Energy transferred to the biphoton as a function of entanglement dimensionality for enhanced ballistic transport and the bound state, respectively (see supplementary material for JSIs corresponding to each dimensionality). In **(E)**, the "rabbit ears" grow as the entanglement dimensionality increases, resulting in a slight increase in standard deviation, as shown in **(D)**. In **(F)**, increasing entanglement dimensionality reduces occurrence of any net energy transport between the modulator and the BFC. Consequently, frequency correlations remain confined to the constant energy axis, i.e., the antidiagonal of the JSI shown in Fig. 2c, for example.

For the biphoton energy bound state described earlier, the situation is slightly different. Here there is a π phase difference between adjacent comb lines, which corresponds to a linear spectral phase shift or simply a time delay. This time delay corresponds to exactly half the modulation period. In other words, the distribution in the relative arrival of signal and idler is now spaced at *half*-integer multiples of the modulation period. Here, unlike in the case of ballistic transport described earlier, photons in an entangled pair experience nearly equal and opposite instantaneous phase shifts [Fig. 3c], which manifests as confinement of frequency correlations to the antidiagonal axis of a JSI measurement. Put another way, the strength of this confinement serves as an indirect probe of the width of any fast substructure in the time correlation function of the BFC. This has major implications for entanglement certification since a direct measurement of the joint temporal correlation is the most straightforward way to establish high-dimensional entanglement using only a white noise model (*32*). Owing to the relatively large timing jitter of SNSPDs (~60-80 ps), fine features in the time-correlation function cannot be resolved directly. While frequency mixing techniques have been developed to overcome this limitation (*26–28*), these measurements become onerous as the dimensionality of the system increases. With increasing attention being paid to wavelength-multiplexed sources for distributing entanglement (*33*), characterizing energy confinement in frequency domain quantum walks offers a potential solution to the challenge of certifying entanglement at nodes or between communication channels in a practical quantum network.

In conclusion, we have demonstrated quantum walks in the frequency domain using two-photon quantum frequency combs that are entangled across multiple dimensions. Furthermore, by adding time to provide a "coin flip" in another degree of freedom (*34, 35*), one can use these quantum walks to implement quantum algorithms such as certification of isomorphism between two high-degree strongly regular graphs (*13*). While our work was carried out using discrete telecom equipment, recent advances in lithium niobate fabrication (*36, 37*) make our platform well-suited to on-chip implementation as well. In fact, the demonstration of ultralow loss integrated devices and high-efficiency nonlinear interactions paves the way for drastically scaling up the dimensions, depth, and circuit programmability of these walks.

**Acknowledgements:** This work was supported in part by the National Science Foundation under award number 1839191-ECCS. The authors thank Paul Kwiat, Ruichao Ma, Christian Reimer, Simeon Bogdanov, and Cristian Cortes for valuable discussions.

**Author contributions:** P. I. developed the idea and designed the experiment. P.I., N.B.L, and M.S.A. performed the experiments with help from D.E.L. P.I., M.S.A., N.B.L, and A.M.W. carried out the theoretical work and analyzed the data. A.M.W. supervised the project.

# Supplementary Material:

- **Experimental setup**

The experimental setup is depicted in Fig. S1. We use a continuous-wave 775 nm laser with about 1 mW power shining on a periodically poled lithium niobite (PPLN) crystal to generate broadband time-frequency entangled photons with about 40 nm (5 THz) bandwidth, with a power of about 5 nW. A pulse shaper is then used to carve this spectrum to make a biphoton frequency comb (BFC) with 25 GHz frequency spacing between the bins and about 9 GHz linewidth. The pulse shaper is also able to manipulate individual frequency bins' phase. After making the high-dimensional entangled state, it is sent to the quantum walk circuit which is a phase modulator driven with a 25 GHz RF sinusoidal waveform. After the quantum walk, another pulse shaper picks two frequency bins at a time and sends them to superconducting single photon detectors (SNSPDs) (Quantum Opus). The relative arrival time of photons on the SNSPD pair is then monitored using an event timer (PicoQuant HydraHarp 400).

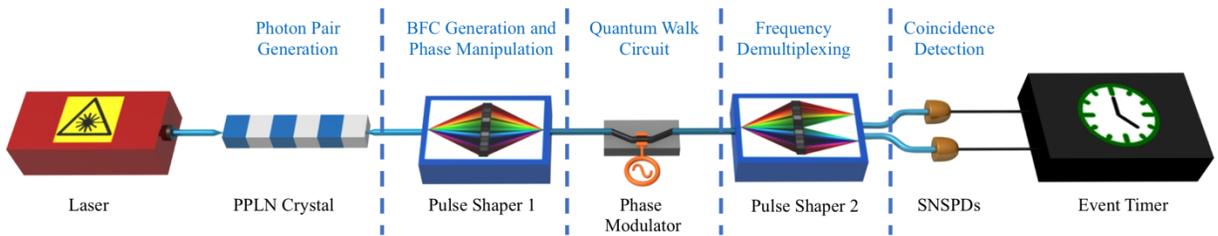

**Fig. S1. The experimental setup.** Broadband time-frequency entangled photon pairs are generated from a continuous-wave laser shining on a PPLN crystal. A BFC is then carved from this continuous spectrum with pulse shaper 1. Pulse shaper 1 can also manipulate the phase of each frequency mode. The high-dimensionally entangled photon pairs then enter the quantum walk circuit, namely a phase modulator driven with a sinusoidal RF waveform whose frequency is identical to the frequency spacing between the adjacent modes of the BFC. After the quantum circuit, two pulse shaper 2 selects two frequency modes at a time and send them to two SNSPDs, where correlations between the two modes are measured.

- **Quantum walk Hamiltonian**

A phase modulator multiplies the wavefunction of an input by $e^{i\delta \cos \omega_m t}$, which can be rewritten as $\sum_{n=-\infty}^{\infty} i^n J_n(\delta) e^{in\omega_m t}$. Therefore, in the Fourier domain, the Hamiltonian of the phase modulation process in terms of creation and annihilation operators can be written as:

$$H = \sum_{n=-\infty}^{\infty} J_n(\delta) \, a^\dagger_{m+n} a_m \quad (1)$$

Which transfers frequency mode $m$ to mode $m + n$ with probability amplitude $J_n(\delta)$, Bessel function of n-th order with modulation depth $\delta$. Given the symmetry equation between positive- and negative-order Bessel functions $J_{-n}(\delta) = (-1)^n J_n(\delta)$, Eq. (1) can be rewritten as:

$$H = J_0(\delta)a_m^\dagger a_m + \sum_{n=1}^{\infty} J_n(\delta)\left(a_{m+n}^\dagger + (-1)^n a_{m-n}^\dagger\right)a_m \quad (2)$$

In our experiments, this Hamiltonian operates on a maximally entangled $d$-dimensional bipartite state of the form:

$$|\psi\rangle_{in} = \frac{1}{\sqrt{d}} \sum_{m=1}^{d} e^{i\theta_m} a_m^\dagger a_{-m}^\dagger \quad (3)$$

Where $\theta_m$ is the phase associated with state of the photon pair in modes $m$ and $-m$. The state at the output of the quantum circuit is then the Hamiltonian acting on both signal and idler photons of the input state, resulting in:

$$|\psi\rangle_{out} = H|\psi\rangle_{in} = \frac{1}{\sqrt{d}} \sum_{m=1}^{d} e^{i\theta_m} \sum_{n,n'=-\infty}^{\infty} J_n(\delta) J_{n'}(\delta) \, a_{m+n}^\dagger a_{-m+n'}^\dagger \quad (4)$$

The probability amplitude of getting a coincidence at the output modes $i$ and $j$, whose absolute value is measured in our joint spectral intensity measurement, can be interpreted with renaming the indices of annihilation operators:

$$|\psi_{j,k}\rangle_{out} = \frac{1}{\sqrt{d}} \sum_{m=1}^{d} e^{i\theta_m} J_{j-m}(\delta) J_{k+m}(\delta) \, a_j^\dagger a_k^\dagger \quad (5)$$

The coincidences measured between modes $i$ and $j$ are:

$$C_{j,k} = \langle \psi_{j,k}|\psi_{j,k}\rangle_{out} = \frac{1}{d}\left|\sum_{m=1}^{d} e^{i\theta_m} J_{j-m}(\delta) J_{k+m}(\delta)\right|^2 \quad (6)$$

Now we consider a couple of special cases. On the antidiagonal terms, where $k = -j$, the coincidences are:

$$C_{j,-j} = \frac{1}{d}\left|\sum_{m=1}^{d} e^{i\theta_m} J_{j-m}(\delta) J_{-j+m}(\delta)\right|^2 \quad (7)$$

And using the symmetry of Bessel functions:

$$C_{j,-j} = \frac{1}{d}\left|\sum_{m=1}^{d} e^{i(\theta_m + \pi(j-m))} J_{j-m}^2(\delta)\right|^2 \quad (8)$$

In the case of the fermionic quantum walk, $\theta_m = m\pi$, which results in:

$$C_{(j,-j)fermionic} = \frac{1}{d}\left|\sum_{m=1}^{d} e^{i\pi j} J_{j-m}^2(\delta)\right|^2 = \frac{1}{d}\left|\sum_{m=1}^{d} J_{j-m}^2(\delta)\right|^2 \quad (9)$$

In Eq. (9), the relative phases between different states drop out, resulting in an always positive coincidence probability in the fermionic quantum walk for the antidiagonal terms, confirming the antibunching process observed in Fig. 2 c.

- **Bosonic and fermionic quantum walks for different modulation depths**

In this section, the theoretical and measured JSI for both bosonic and fermionic quantum walks are shown. The experimental data are in good agreement with the simulated JSIs.

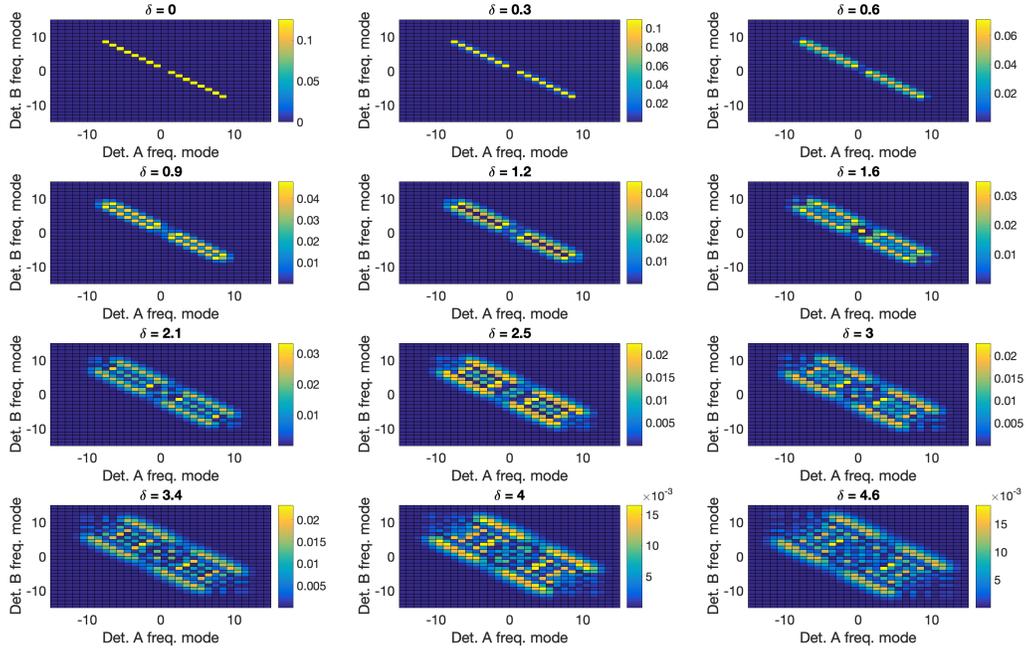

**Fig. S2. Bosonic quantum walk (simulated).** Theoretical JSI of bosonic quantum walk for an eight-dimensional entangled state as a function of modulation depth. Since the upper-left triangular half of the JSI is identical to the lower-right triangular half, the total number of coincidences in the matrix is normalized to 2.

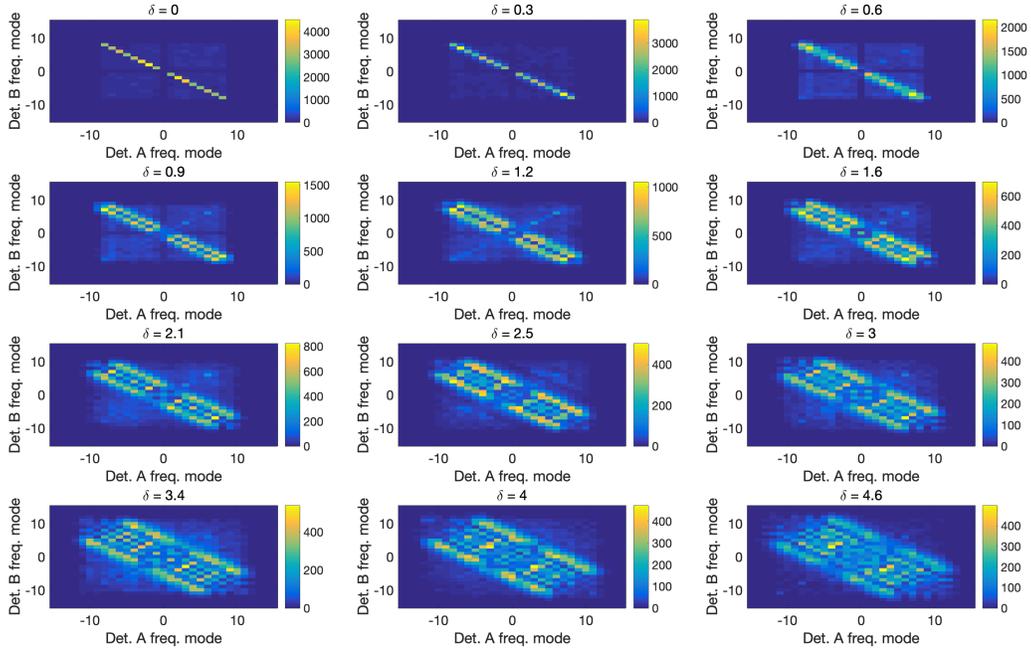

**Fig. S3. Bosonic quantum walk (experiment).** Experimental JSI of bosonic quantum walk for an eight-dimensional entangled state as a function of modulation depth.

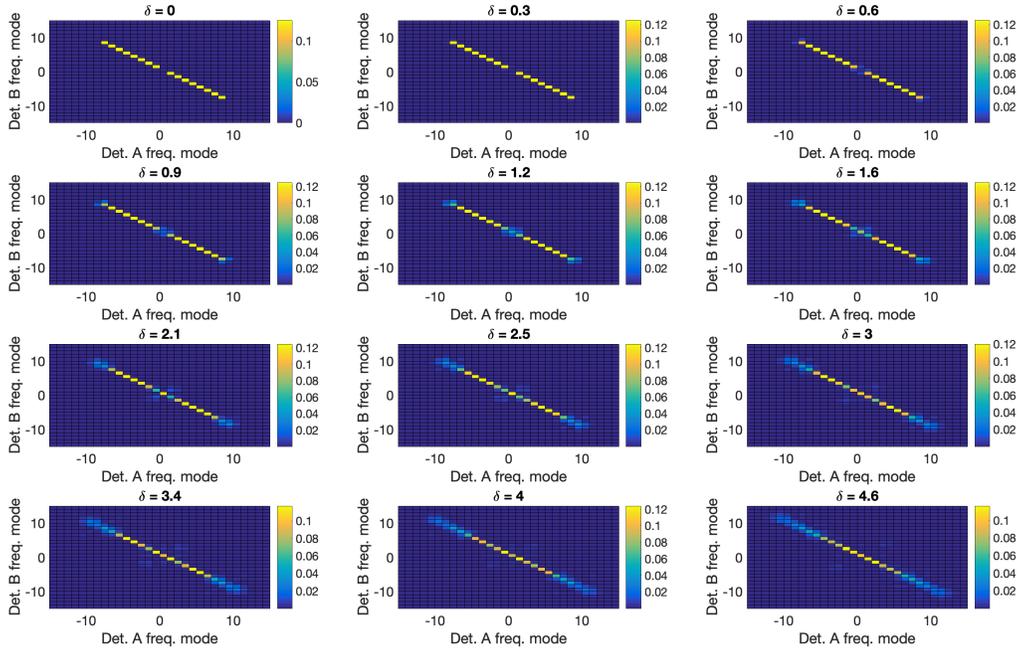

**Fig. S4. Fermionic quantum walk (simulated).** Theoretical JSI of fermionic quantum walk for an eight-dimensional entangled state as a function of modulation depth.

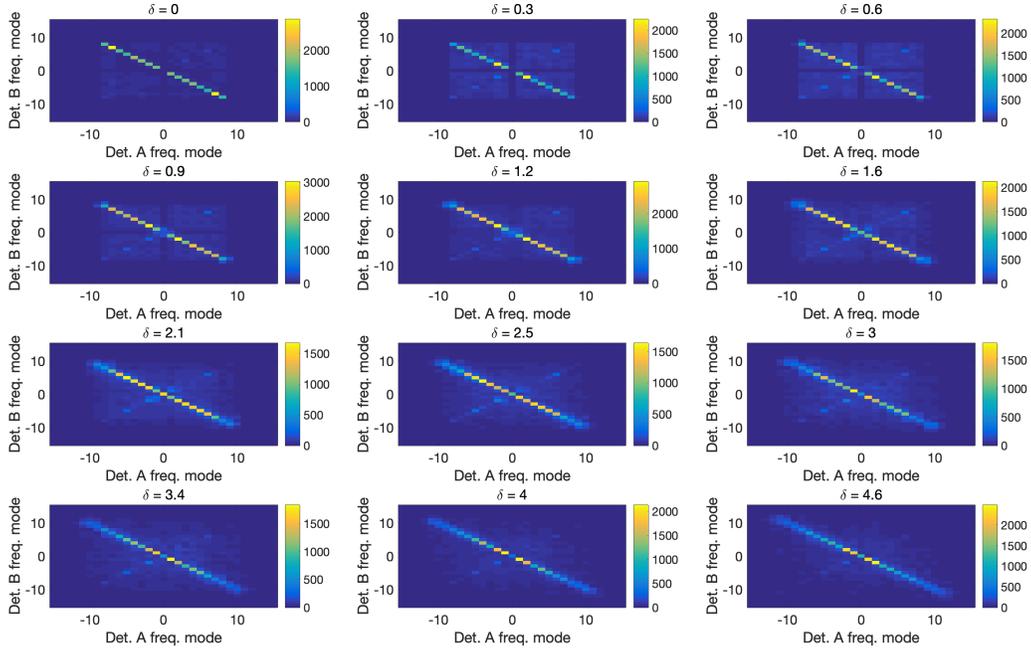

**Fig. S5. Fermionic quantum walk (experiment).** Experimental JSI of fermionic quantum walk for an eight-dimensional entangled state as a function of modulation depth.

- **Incoherent quantum walk**

Here, we show the simulated and experimental data for the incoherent quantum walk. To retrieve the JSI of an incoherent correlated state, we pick two correlated bin pairs $|m,-m\rangle_{SI}$ for $m = 1,…,8$ at a time and measure the output JSI. We then add these JSIs incoherently, which in equivalent to adding the probabilities of each output state incoherently not adding probability amplitudes (Fig. S8). Note that sharp structure evident in quantum walks of photons corresponding to coherent superpositions of frequency bins is no longer observed.

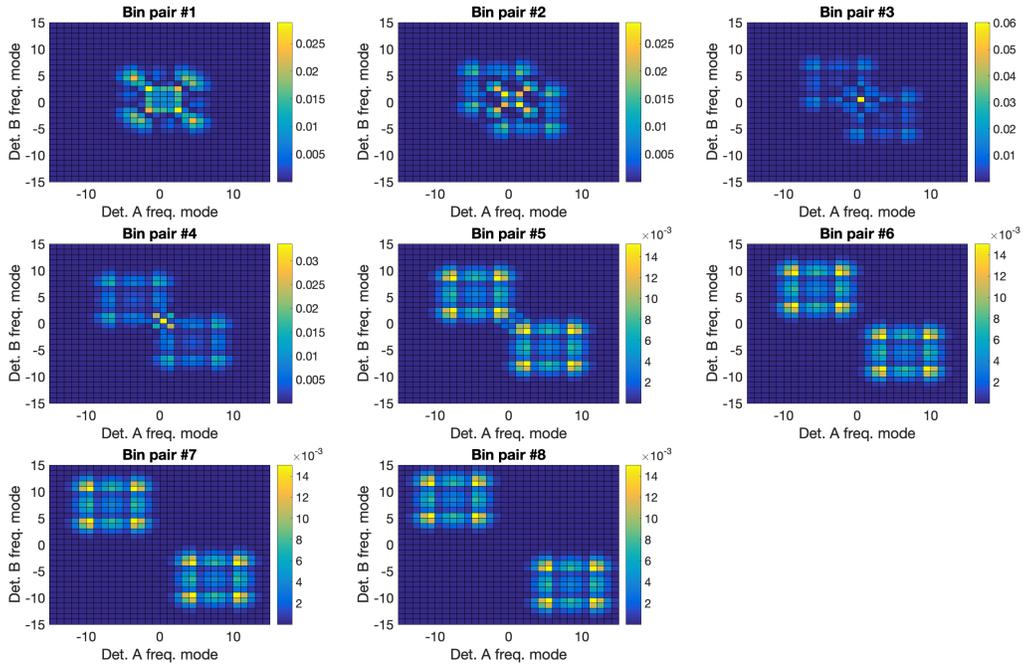

**Fig. S6. Output JSI for individual bin pairs (simulated).** Theoretical JSIs corresponding to the input states $|\psi\rangle = |m, -m\rangle_{\text{SI}}$ for $m = 1, \ldots, 8$ after the quantum walk circuit.

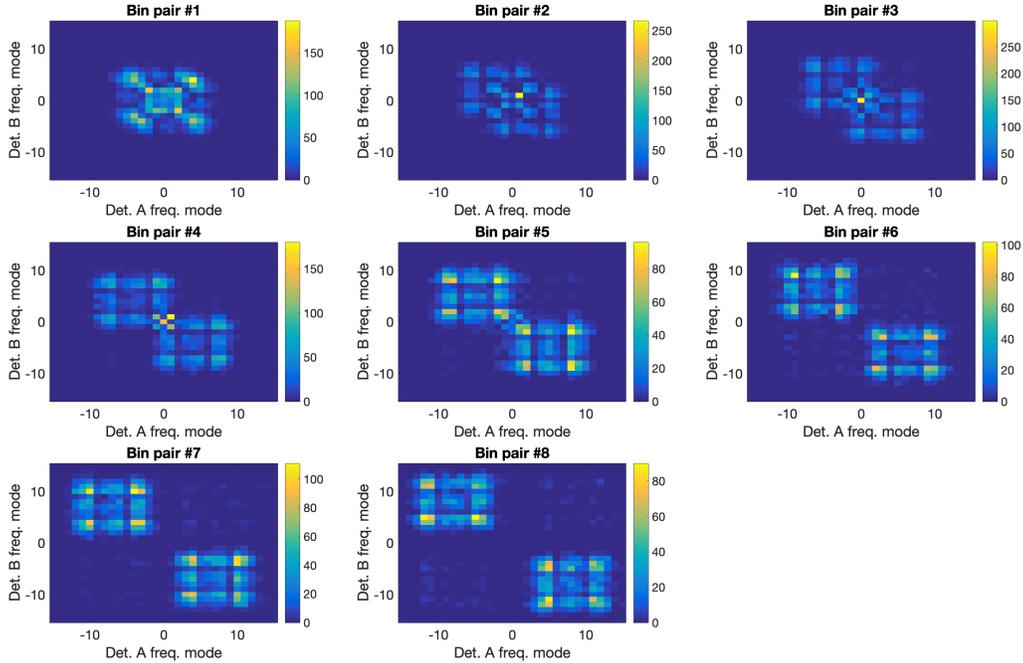

**Fig. S7. Output JSI for individual bin pairs (Experiment).** Experimental JSIs corresponding to the input states $|\psi\rangle = |m, -m\rangle_{\text{SI}}$ for $m = 1, \ldots, 8$ after the quantum walk circuit.

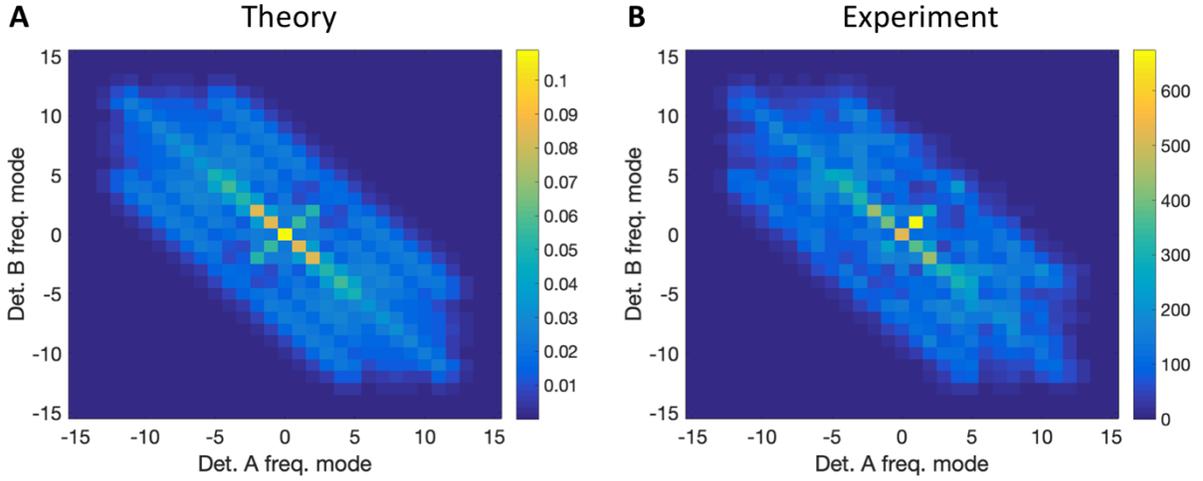

**Fig. S8. Adding the output JSIs of states $|m, -m\rangle_{SI}$ for $m = 1, ..., 8$.** The addition of the results is equivalent to the output of an incoherent mixture of correlated frequency pairs. The experimental results (**A**) match with theoretical simulations (**B**).

- **Transferred energy between the quantum walk circuit and the biphotons as a function of entanglement dimensionality**

Here, we show the transferred energy between the phase modulator and the biphotons for both enhanced ballistic transport (Fig. S9) and biphoton energy bound state (Fig. S10), for fixed modulation depth of $\delta = 6.1$. In this section, based on availability of equipment, we used more modulation depth to see the features of varying entanglement dimensionality more clearly.

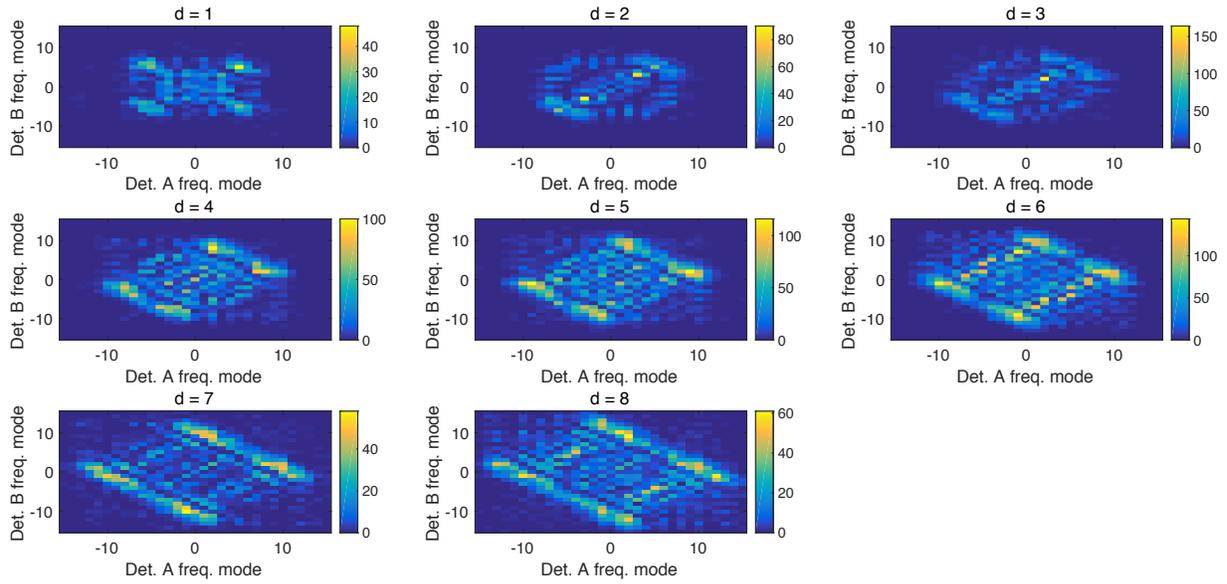

**Fig. S9. JSI enhanced ballistic energy transport as a function of entanglement dimensionality (experiment).** Experimental JSI of bosonic quantum walk for a fixed modulation depth ($\delta = 6.1$) as a function of entanglement dimensionality.

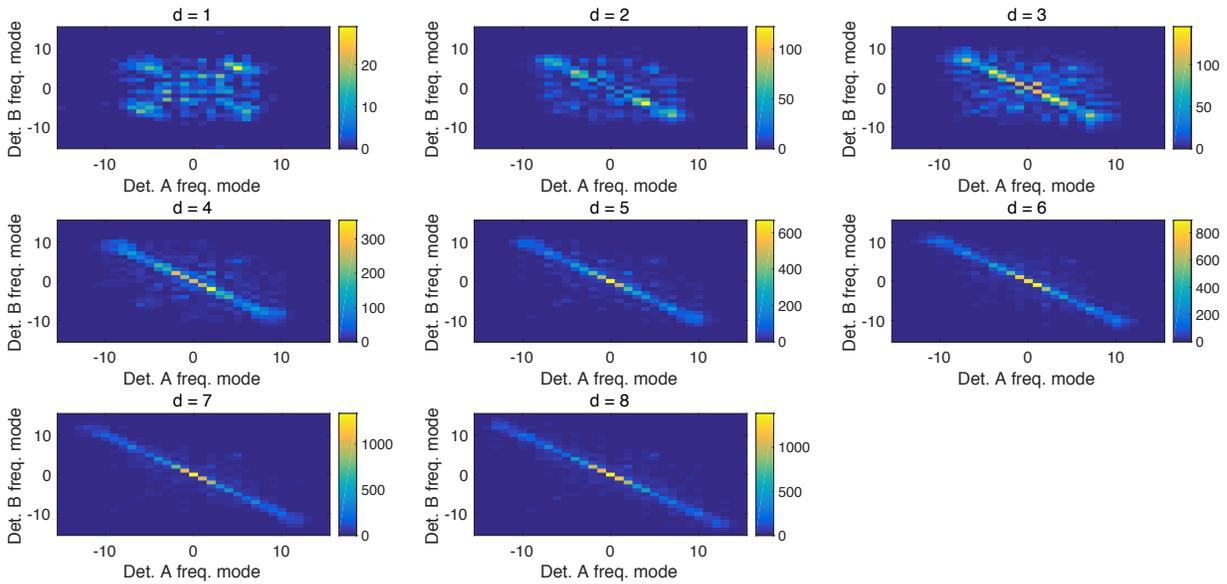

**Fig. S10. JSI of the energy bound state as a function of entanglement dimensionality (experiment).** Experimental JSI of fermionic quantum walk for a fixed modulation depth ($\delta = 6.1$) as a function of entanglement dimensionality.

- **Quantum walk in the limit of high modulation depth**

The distinction between the quantum walk distribution of the bosonic, fermionic and the incoherent case becomes more evident when looking at the JSI (Fig. S11) in the limit of high modulation depth ($\delta=200$). Bosons spread diagonally while fermions antibunch and spread antidiagonally. Furthermore, the ability to modify the initial state allows simulating anyonic behavior by setting the linear spectral phase to having a slope of $\pi/2$, i.e. $|\psi\rangle = 1/\sqrt{8} \sum_{m=1}^{8} e^{im\pi/2}|m,-m\rangle_{SI}$,

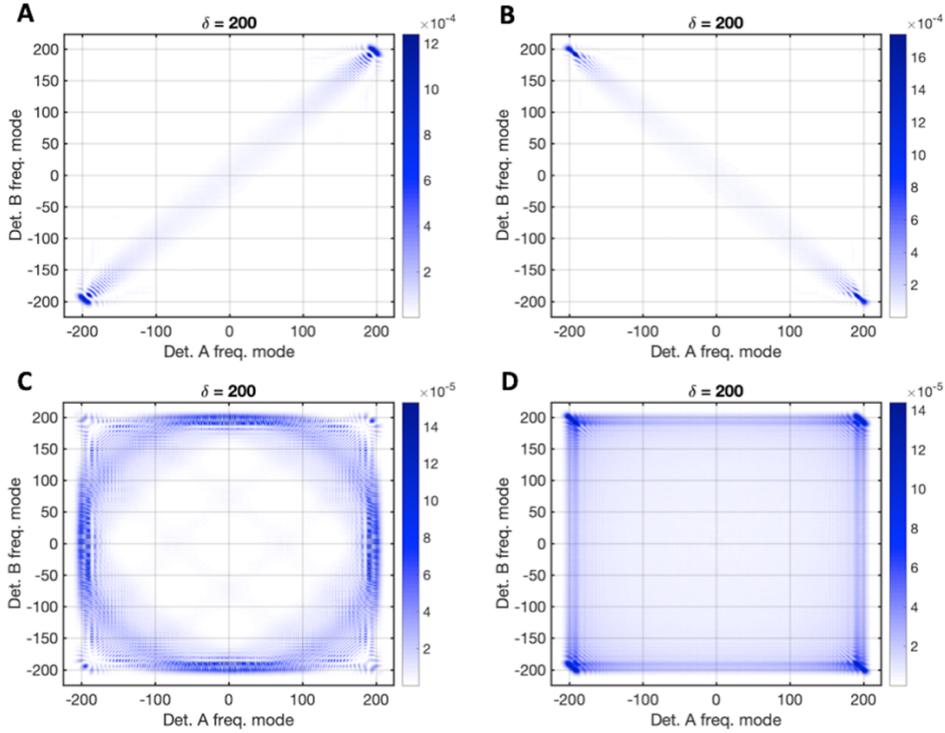

**Fig. S11. Output JSI in the limit of high modulation depth ($\delta=200$).** (**A**) Bosonic, (**B**) Fermionic, (**C**) Anyonic and (**D**) incoherent case, all simulated. (**A**) and (**B**) in the limit of high modulation clearly show bunching and antibunching effects, respectively, whereas in (**C**) the biphotons relative delay is *quarter*-integer multiple of the modulation period resulting in a quasi-circular pattern. In (**D**), the photons are less correlated with each other.